\documentclass[aps,twocolumn,showpacs,superscriptaddress,groupedaddress]{revtex4}  
\usepackage{graphicx}  
\usepackage{dcolumn}   
\usepackage{bm}        
\usepackage{amssymb, amsmath, amsfonts}   
\usepackage{lipsum}        
\usepackage{color}
\usepackage{comment}

\DeclareGraphicsExtensions{{.pdf}}
\graphicspath{{./}{./figure}}

\begin{document}
\newcommand{\note}[1]{{\color{red}{#1}}}
\newcommand{\noter}[1]{{\color{red}{#1}}}
\newcommand{\noteb}[1]{{\color{blue}{#1}}}

\widetext


\title{
Hydrodynamic alignment of microswimmers in pipes 
}
\author{Norihiro Oyama}
\affiliation{Mathematics for Advanced Materials-OIL, AIST-Tohoku University, Sendai 980-8577, Japan}
\author{John Jairo Molina}
\affiliation{Department of Chemical Engineering, Kyoto University, Kyoto 615-8510, Japan}
\author{Ryoichi Yamamoto}
\affiliation{Department of Chemical Engineering, Kyoto University, Kyoto 615-8510, Japan}
\affiliation{Institute of Industrial Science, The University of Tokyo, Tokyo 153-8505, Japan.}
\date{\today}

\begin{abstract}
  We investigate the dynamics of model microswimmers under
  confinement, in cylindrical geometries, by means of three dimensional direct numerical calculations
  with fully resolved hydrodynamics.
  Such swimmers are known to show collective alignment in bulk,
  and we confirmed that similar alignment can be observed even in pipes,
  although the volume fraction dependency shows qualitative differences.
  By comparing the structural information, we investigated the cause of
  such differences. We found an order/disorder
  phase transition for a specific type of swimmers,
  as the size of the pipe becomes comparable to the size of the swimmers.
  Such dynamics are found not to depend on the geometry of confinement.
\end{abstract}

\maketitle

\setcounter{section}{0}
\section{\label{sec:level1}Introduction}
Self-propelled particles (SPPs)
are attracting more and more interest as
a representative example of out-of-equilibrium systems\cite{Marchetti:2013tq,Takatori:2014do,Solon:2015bt,Fodor:2016fm,Nikola:2016jc,Zottl:2016kr,Bechinger:2016cf,Speck:2016jl,Blaschke2016,Winkler2016,Lauga2009,Elgeti2015}.
Examples of SPPs range from the microscopic scale, with algae and bacteria, to
the macroscopic scale, which includes all animals, and recently,
even artificial SPPs have been constructed, such as active Janus particles\cite{Paxton:2004ea,Paxton:2006dm,Popescu2016,Kroy2016} or
self-propelled droplets\cite{Seemann2016,Izri:2014}.
Among them, microswimmers are of particular interest\cite{Elgeti2015},
since they are suited to well-controlled lab
experiments\cite{Lushi:2014fn}, and have many potential applications,
for example, as targeted drug delivery systems.
It is known that nontrivial motion, like 
the collective alignment of the swimming direction or the dynamic
clustering can be observed,
even for systems where particles interact with each other
only through hydrodynamic interactions and excluded volume effects\cite{Evans:2011cw,Alarcon:2013dg,Oyama:2016eh}.
Such collective motion is mainly due to the complicated hydrodynamic interactions,
and cannot be predicted only from knowledge of the single particle
dynamics. That these hydrodynamic interactions are strongly affected
by the presence of confining walls is well known\cite{Elgeti2016,Wysocki2015,Spagnolie2015}. 
Experimentally, Das {\it et al.} reported that when a single Janus particle
swims in the vicinity of a wall, it tends to swim along the wall\cite{Das:2015ef}.
Lushi {\it et al.} investigated the collective dynamics
of a bacterial dispersion inside a circular confinement,
and reported that the system spontaneously shows an anomalous
double-vortex motion\cite{Lushi:2014fn}.
Regarding the theoretical or numerical studies with full hydrodynamics, while most
of them have focused on the dynamics in bulk, 
several works have reported on the dynamics of
microswimmers near walls or under
confinement\cite{Lushi:2014fn,Das:2015ef,Zhu:2013js,Li:2014bu,Zottl:2014wy,Oyama:2016eh,Ishimoto2013,Theers2016,Spagnolie2012,Malgaretti2017}
  The dynamics under confinement is usually studied using flat parallel
  walls or cylindrical pipes.
  We would like to note that works with nonspherical swimmers\cite{deGraaf:2016fc}
  or chemically actuated swimmers\cite{Popescu:2009it} with confinements
  have been reported.
  Also, the dynamics of microswimmers in external field\cite{Kuhr2017,Stark2016a} or with background flow\cite{Zottl2012,Jibuti2014a} have also been investigated.
In the present work, we focus on spherical swimmers which self-propel by utilizing
only the hydrodynamic interactions between the ambient fluid,
so-called ``squirmers''\cite{Lighthill:1952ta,Blake:2006ig,Ishikawa:2006hf,Blaschke2016}.
In this description of microswimmers, different types of
swimming mechanisms can be expressed only by changing one parameter,
which we refer to as the swimming parameter.
As examples of the work using the same swimmer description as
ours,
Li and Ardekani have investigated the static structure
between flat parallel walls
and shown evidence for the accumulation of particles near the wall\cite{Li:2014bu}.
Z\"ottl and Stark have studied a similar system\cite{Zottl:2014wy},
but under extreme confinement, and observed a dense-dilute phase
separation, which is not seen in bulk.
Our previous work\cite{Oyama:2016eh} has focused on the dynamic properties of a microswimmer dispersion 
confined between flat parallel walls with relatively large separations
(compared to the size of the particles).
In ref.~\cite{Oyama:2016eh}, we observed a traveling wave-like
collective motion for a specific range of swimming parameters and densities.
Though such a motion would seem to be a consequence of confinement,
we clarified that it can be understood as the manifestation of the
pseudo-acoustic properties of the system, which is already observed in bulk.
As shown here, even for the same confinement geometry,
the dynamic behavior can be considerably different depending on the strength
of the confinement (i.e., the wall separation). Therefore, how the
dynamics changes under confinement is a very difficult question to answer.

Although we can find several works on the collective dynamics
in bulk or between flat parallel walls,
the many particle dynamics of swimmers in pipes has not been extensively studied
so far. This is the focus of the present work.
Intuitively, we can expect that for big enough pipes, the dynamics
will be the same as for bulk systems. Therefore, we focus on the dynamics in pipes
with diameters comparable to that of the particle, where we can expect
nontrivial behaviors which are different from those in bulk. 
In fact, in the work by Zhu {\it et al.}, which deals with the single
particle dynamics in a pipe,
it is reported that different dynamical modes can be
observed depending on the swimming type and strength\cite{Zhu:2013js}.
Taking into account the possibility that the size and the shape of the
pipe affect the dynamic properties, in this work, we investigated the
collective alignment effects, known as the polar order formation,
varying the four main parameters,
namely, the pipe size,
the pipe shape, the volume fraction and the type of the swimmers.
Regarding the polar order formation in many particle
systems, the behavior in pipes is mostly the same as in bulk, if the pipe size is large enough.
However, for a specified region of the parameter space, we observed clear wall
effects, which we investigated by measuring the structural information
of the dispersion.
In addition, we observed the pipe size dependent order/disorder phase transition only for
the parameters at which clustering behaviors have been reported
in bulk and between flat walls\cite{Alarcon:2013dg,Oyama:2016eh}.
In this work, we also measured the bulk structural
information and obtained indirect evidences which states
that the clustering is important for the collective alignment for a
specific range of parameters.

\section{\label{sec:level2}Simulations}
\subsection{\label{sec:level21}The Squirmer Model}
As the numerical model for microswimmers, 
we employed the squirmer model\cite{Lighthill:1952ta,Blake:2006ig}.
In this model, the microswimmers are expressed by rigid particles
with a prescribed flow field on their surface.
The general squirmer model is expressed in the form of an infinite
expansion, with components along the tangential, radial, and azimuthal directions.
However, utilizing only the first two modes of the tangential field,
following Eq.~(\ref{eq:Sq_2}),
already enables us to model different types of swimmers, namely pushers,
pullers and the neutral swimmers.
\begin{eqnarray}
\boldsymbol{u}^s(\theta) = B_1\left(\sin{\theta} + \frac{\alpha}{2}\sin{2\theta}\right)\hat{\boldsymbol{\theta}}
\label{eq:Sq_2},
\end{eqnarray}
where, $\boldsymbol{u}^s$ denotes the surface flow field,
$\hat{\boldsymbol{r}}$ is a unit vector directed from the center of the
particle to a point on its surface,
$\theta=\text{cos}^{-1}\left(
\hat{\boldsymbol{r}}\cdot\hat{\boldsymbol{e}}\right)$ the polar
angle between
$\hat{\boldsymbol{r}}$ and
the swimming direction $\hat{\boldsymbol{e}}$,
and $\hat{\boldsymbol{\theta}}$ is the tangential
unit vector at $\hat{\boldsymbol{r}}$.
This simplified squirmer model has been widely used and is known to
lead to a wide variety of nontrivial phenomena\cite{Zottl:2016kr}.
The coefficient of the first mode, $B_1$, determines the swimming
velocity of an isolated squirmer ( $U_0=2/3B_1$), and
that of the second mode, $B_2$, determines the stresslet\cite{Ishikawa:2006hf}.
The ratio between the coefficients of the two modes, $\alpha=B_2/B_1$
in Eq.~(\ref{eq:Sq_2}),
determines the swimming type and strength.
In the following, we call $\alpha$ the swimming parameter.
Negative values of $\alpha$ represents pusher-type swimmers, which
swim with an extensile flow field in the swimming direction,
and positive values describe
puller-type swimmers, which swim with a contractile flow.
A value of $\alpha=0$ stands for a neutral swimmer, which
generates a potential flow field.
In this model, the prescribed flow field is assumed to be
an axisymmetric pure tangential one and the particle is rigid and spherical.
These are the only assumptions employed to derive the model.
Therefore, it is thought that this model can capture even
the features of the artificial microswimmers like Janus particles
even though it was originally intended to describe ciliary propulsion
of micro-organisms.

\subsection{\label{sec:level3}Smoothed Profile Method (SPM)}
In this work, we investigate the dynamics of squirmer particles swimming in a viscous
fluid.
For this, we must solve the combined fluid-solid problem.
The Newton-Euler equations of motion govern the particle trajectories:
\begin{align}
  \dot{\boldsymbol{R}}_i &= \boldsymbol{V}_i 
  &\dot{\boldsymbol{Q}}_i &= \text{skew}(\boldsymbol{\Omega}_i)\cdot
  \boldsymbol{Q}_i\label{Eq:NE}\\
  M_{\text{p}}\dot{\boldsymbol{V}}_i &= \boldsymbol{F}_i^{\text{H}}+\boldsymbol{F}_i^{\text{C}}
  &\boldsymbol{I}_{\text{p}}\cdot\dot{\boldsymbol{\Omega}}_i &= \boldsymbol{N}_i^{\text{H}}\notag
\end{align}
where the subscript $i$ means the particle index,
$\boldsymbol{R}$ the position vector,
$\boldsymbol{Q}$ the orientation matrix,
$\text{skew}(\boldsymbol{\Omega})$ the skew symmetric matrix of
the angular velocity
$\boldsymbol{\Omega}$, and $\boldsymbol{F}^{\text{H}}$ and
$\boldsymbol{N}^{\text{H}}$ the hydrodynamic force
and torque though which the particle dynamics is coupled with the flow field.
The interaction between particles, $ \boldsymbol{F}^{\text{C}} $, is
also considered to prevent overlapping of the particles
by employing a repulsive Lennard-Jones type potential,
or the Weeks-Chandler-Andersen potential.
As the powers for the potential, we adopted 36-18:
\begin{align}
  \boldsymbol{F}_i^{\rm C} &= \sum_j\boldsymbol{F}_{ij},
  \boldsymbol{F}_{ij} = -\boldsymbol{\nabla}_i U\left(r_{ij} \right),\label{Eq:Fc}\\
  U\left( r_{ij}\right) &=
  \begin{cases}
    4\epsilon\left[ \left( \frac{\sigma}{r_{ij}}\right)^{36} -
      \left( \frac{\sigma}{r_{ij}}\right)^{18}\right] + \epsilon
    &\left( r_{ij}\le  r_{\rm C}\right) \\
    0 & \left( r_{ij}\ge r_{\rm C}\right)
    \end{cases}
\end{align}
where, $r_{\rm C} = 2^{1/18}\sigma $ is the cutoff length. 
$\epsilon$ gives the strength of the potential and is the unit of energy in the system.
Due to the steep nature of the potential,
the value of $\epsilon$ affect only weakly to the system dynamics.
The time evolution of the fluid flow field $\boldsymbol{u}_{\rm f}$ is described by the
Navier-Stokes equation with the incompressible condition:
   \begin{align} 
	\boldsymbol{\nabla}\cdot\boldsymbol{u}_{\rm f} &= 0\\
                \rho_{\text{f}} \left( \partial_t + \boldsymbol{u}_{\text{f}}\cdot\nabla\right)\boldsymbol{u}_{\text{f}}
                &= \nabla \cdot \boldsymbol{\sigma}_{\rm f}\\
                \boldsymbol{\sigma}_{\rm f}&=-p\boldsymbol{I}+\eta \left\{ \nabla\boldsymbol{u}_{\text{f}} + \left( \nabla\boldsymbol{u}_{\text{f}} \right)^t \right\}
   \end{align}
where $\boldsymbol{\sigma}_{\rm f}$ is the stress tensor,
$\rho_\text{f}$ the mass density of the host fluid and  $\eta$ the shear viscosity.
In order to solve these simultaneous equations efficiently, we employed the Smoothed
Profile Method (SPM).
In the SPM, we don't treat the sharp boundary between the
solid and fluid phases explicitly.
Instead, a diffuse interface with thickness of $\xi$ is introduced,
and the phase boundaries are expressed by using a continuous order
parameter $\phi$, which takes the value of one in the solid domain,
zero in the fluid domain, and intermediate values within the diffuse
interface region.
Employing this method, all the physical quantities can be expressed
in terms of continuous fields defined over the whole computational
domain.
The total velocity field $\boldsymbol{u}$ which includes both particle and
fluid velocity information can be expressed like:
\begin{align}
  \boldsymbol{u} &= \left( 1- \phi\right)\boldsymbol{u}_{\rm f}
  + \phi \boldsymbol{u}_{\rm p},\notag\\
  \phi\boldsymbol{u}_{\rm p} &= \sum_i\phi_i\left[
    \boldsymbol{V}_i+\boldsymbol{\Omega}_i\times\boldsymbol{R}_i\right]
\end{align}
where, $\left(1-\phi\right)\boldsymbol{u}_{\rm f}$ and
$\phi\boldsymbol{u}_{\rm p}$ are the contribution of the fluid flow
field and the particle motion to the total velocity field.
And now, the time evolution of the total velocity field
$\boldsymbol{u}$
is governed by a modified incompressible Navier-Stokes equation:
\begin{align}
  \rho_{\rm f}\left(
  \partial_t+\boldsymbol{u}\cdot\boldsymbol{\nabla}
  \right)\boldsymbol{u}&=
  \boldsymbol{\nabla}\cdot\boldsymbol{\sigma}_{\rm f}+\rho_{\rm f}
  \left(\phi\boldsymbol{f}_{\rm p}+\boldsymbol{f}_{\rm sq}\right),\\
  \boldsymbol{\nabla}\cdot\boldsymbol{u}&=0,
\end{align}
where $\phi\boldsymbol{f}_{\text{p}}$ is
the body force guaranteeing the rigidity condition of particles and
$\boldsymbol{f}_{\text{sq}}$ the force needed to maintain the
squirming motion.
By doing this, the calculation cost can be
reduced by orders of magnitude. See refs.\cite{Nakayama2005a,Kim2006a,Nakayama:2008fi,Molina:2013hq} for more
detailed information about this method.

In this work, the dynamics of swimmers in pipes are considered.
For this, we have to model solid walls numerically.
Within the SPM scheme,
the walls can be expressed by assemblies of
particles which are pinned and not allowed to translate or rotate.
By this, the stick boundary at the wall surface is guaranteed.
The wall particles are placed uniformly and
have the same diameter as the swimmer particles.
Because we consider various shapes and sizes of pipes,
the number of particles composing the walls varies depending on the pipe.
  We confirmed that the roughness
  due to the wall representation as a collection of discrete particles
  has only a weak effect on the particle dynamics.
  All the quantities are non-dimensionalized in terms of the grid spacing $\Delta$,
  the viscosity $\eta$ and the density of the fluid $\rho_{\rm f}$ in our calculations and these
  values are set to be unity.
  No noise is included in any of the calculations in this work (the limit of the infinite Peclet number), unless stated otherwise.

\subsection{System Parameters}
Using the calculation method presented above, we conducted three-dimensional
direct numerical simulations (DNS) of microswimmer dispersions
confined in pipes.
The diameter $\sigma$ and interface thickness $\xi$ of particles are set
to be $\sigma=6\Delta$ and $\xi=2\Delta$ respectively, where
$\Delta$ stands for the grid spacing.
The excluded volume effect between particles (both wall-particle and
particle-particle) is implemented by $\boldsymbol{F}^{\rm C}$ in Eq.~(\ref{Eq:NE}),(\ref{Eq:Fc}).
Regarding the parameters in Eq.~(\ref{eq:Sq_2}), 
we set $B_1=0.25$ for all the
calculations and varied $\alpha$ to study the $\alpha$ dependency of
the dynamics.
This value of $B_1$ gives a particle Reynolds number of one for an
isolated swimmer~(as presented above, only the value of $B_1$
determines the steady state velocity).
The symmetry axis of the pipe is defined to be parallel to the $y$ axis.
In what follows, we refer to the confinement with a circular cross
section as a pipe,
and the one with a rectangular cross section as a duct.
Regardless of the size or shape of confinement, the length of the pipe $L$
is set to be $L=128\Delta\approx 20\sigma$
and the periodic boundary condition is
set in this direction (in other directions, the system is enclosed by
the confinement, and the system size in those directions depends on
the confinements).
Unless stated otherwise, thermal fluctuations are ignored throughout this paper.
We will consider the pipe with
diameter $D=8\sigma$ to be our reference confinement system, where $D$ is defined as the diameter of the pipe
containing the free space available to the particle.
In this article, first we present the dynamics of this
reference system, and then those obtained for different pipe sizes.
For the definition of the particle volume fraction we used:
\begin{align}
  \varphi = \cfrac{N_{\rm p} V_{\rm p}}{V_{\rm M}},
\end{align}
where $V_{\rm p}=\frac{1}{6}\pi {\sigma}^3$ is the volume of
one particle and $V_{\rm M}$ the volume in which particle centers can move freely:
\begin{align}
  V_{\rm M} = \frac{\pi}{4} L D^2.
  \label{eq:VM}
\end{align}

\begin{figure}
 \begin{center}
  \includegraphics[width=\linewidth]{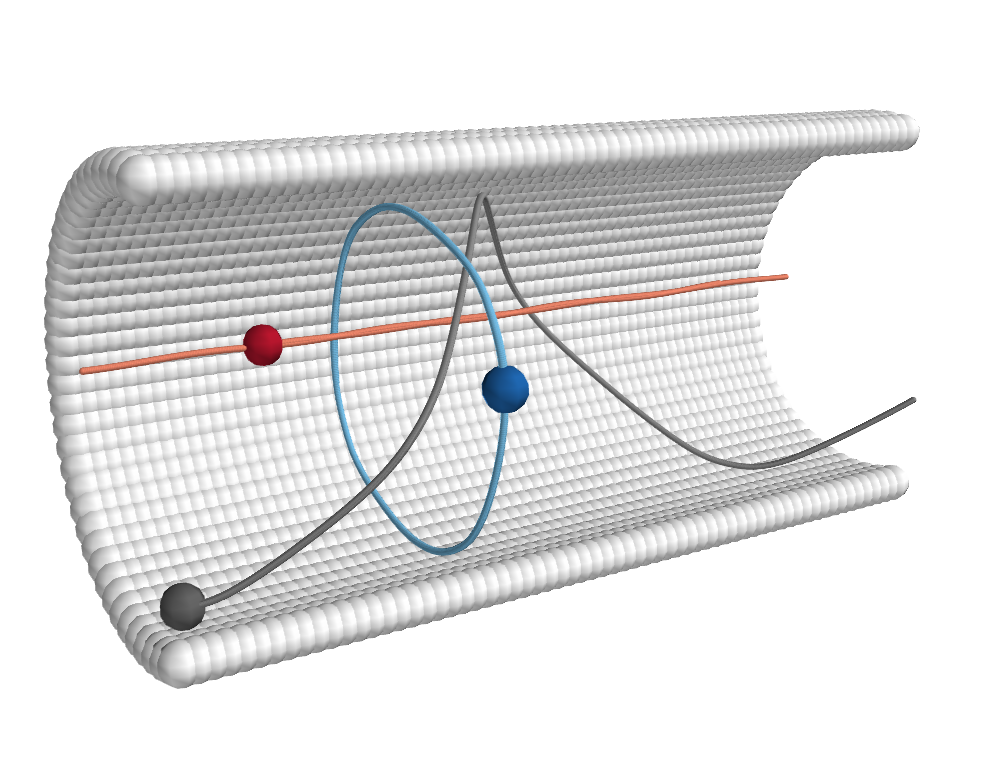}
  \caption{\label{fig:trajectories}
  Typical trajectories of the three different dynamical
  modes. (a) Pusher's rotational orbit ($\alpha=-1$; represented by blue)
  (b) Neutral swimmer's spiral trajectory ($\alpha=0$; represented by gray)
  (c) Puller's rectilinear path ($\alpha=1$; represented by red).
  Note that these three trajectories are obtained using the same
  initial position and orientation.
  Only the trajectories in steady states are shown.
  White particles represent the particles composing the pipe.
  To facilitate the visualization, only half of the pipe-particles are
  drawn.
}
 \end{center}
\end{figure}

\section{\label{sec:level4}Results \& Discussions}
\subsection{Single-particle Dynamics in a Pipe}
The single-particle dynamics in a pipe has been studied by
Zhu {\it et al.}, who reported that
distinct dynamical modes are observed depending on the
swimming parameter $\alpha$\cite{Zhu:2013js}.
In order to confirm the qualitative agreement within
our calculation scheme,
we performed calculations with one particle for various values of
$\alpha$.
As mentioned above, in this section,
we consider only a pipe with diameter $D=8\sigma$, the reference system.
After a sufficiently long time, all the cases achieve steady states.
Typical trajectories of the steady states are shown in
Fig.~(\ref{fig:trajectories}).
Here, the result of the cases with $\alpha=0,\pm 1$ are shown.
The dynamical mode at the steady state depends mostly on the sign of
$\alpha$:
a pusher gets trapped in an orbit and shows constant rotation at a
fixed $y$ position, while a puller
shows rectilinear translation along the pipe axis
at a constant distance from the wall.
A swimmer with $|\alpha| \le 0.4$,
including a neutral swimmer, achieves the combined dynamics of the  
two modes: spiral motion along the pipe axis (rotation + translation).
The pitch of the spiral depends on the value of $|\alpha|$ and
decreases (increases) with the increase of $|\alpha|$ for the case of a pusher (puller).
And eventually, at around $|\alpha|=0.4$, the dynamical mode changes.
These results are consistent with the results in
ref.\cite{Zhu:2013js}; although the trajectories of the pushers are
not analyzed in detail in ref.\cite{Zhu:2013js}.

As shown here, the single-particle dynamics are affected by the
confinement.
The intriguing thing is that the dynamical mode
depends on the swimming parameter $\alpha$.
Regarding the bulk collective properties, it is known that such a swimming parameter
dependence exists as well.
For example, the polar order $P$ (Eq.~(\ref{eq:bulk_Q})), often used to measure the
degree of collective alignment, strongly depends on $\alpha$.
\begin{align}
  P= \left< \frac{1}{N_{\rm p}}\left| \sum_i^{N_{\rm p}} \hat{\boldsymbol{e}}_i\right|\right>,
  \label{eq:bulk_Q}
\end{align}
where, $N_{\rm p}$ stands for the number of particles in the system
and $\hat{\boldsymbol{e}}_i$ the unit swimming direction vector of
particle $i$.
This quantitative measure of the degree of order, $P$,
takes a value of one when the system is completely aligned,
and $P\approx P_0= 1/\sqrt{N_{\rm p}}$ when the system is completely isotropic.
The value of the polar order $P$ decreases as the absolute value of $\alpha$
increases.
  Though the dynamical mode transition of the single swimmer motion in pipe occurs
  at the same absolute value of $\alpha$, it is known that the bulk polar order $P$ is
  asymmetric with respect to $\alpha$: the order is broken faster for pushers than for pullers
  with increasing $|\alpha|$\cite{Oyama:2016eh,Alarcon:2013dg,Evans:2011cw}.

\subsection{Many Particle Dynamics in a Pipe}
We next performed calculations for many particle systems in a pipe of diameter $D=8\sigma$.
The polar order $P$ is a function of both the swimming
parameter $\alpha$, and the
volume fraction of the particles $\varphi$.
First of all, let us recapitulate the behaviors
in bulk systems\cite{Oyama:2016vj}.
If we measure the polar order as a function of the swimming parameter $\alpha$ at a certain
value of the volume fraction,
the polar order $P$ decreases with the increase in the absolute value of
$\alpha$, as stated above.
This tendency is asymmetry with respect to positive and negative
values of swimming parameter $\alpha$:
with negative values (pushers), showing a faster decay.
If we regard the order parameter $P$ as a function of the particle volume fraction $\varphi$
for a certain value of the swimming parameter $\alpha$,
the behavior is most easily understood in terms of the deviations from
the volume fraction dependence for neutral swimmers.
This reference system ($\alpha=0$) shows two regimes:
for $\varphi \le 0.4$, the values are constant;
for $\varphi \ge 0.4$ there is an abrupt drop
in the order parameter to that of a disordered state $P\approx P_0$.
A similar behavior is observed for pushers (negative values of the swimming parameter
$\alpha$).
At small volume fractions, they show only a weak decrease with
increasing volume fraction, and then exhibit an abrupt drop at
a critical volume fraction $\varphi_c$.
Pushers show a decrease both in the degree of ordering and the
value of $\varphi_c$.
Pullers, on the other hand, show a decrease only in the degree of ordering,
and systems of intermediate pullers~($\alpha=0.5$, for example) even maintain the order at very high
volume fractions, where all other systems show $P\approx P_0$.
Specifically, we stress that in bulk, the system with $\alpha=1$ shows a nonzero
value of the polar order $P$ up to very high volume fractions.

\begin{figure}
  \includegraphics[width=\linewidth]{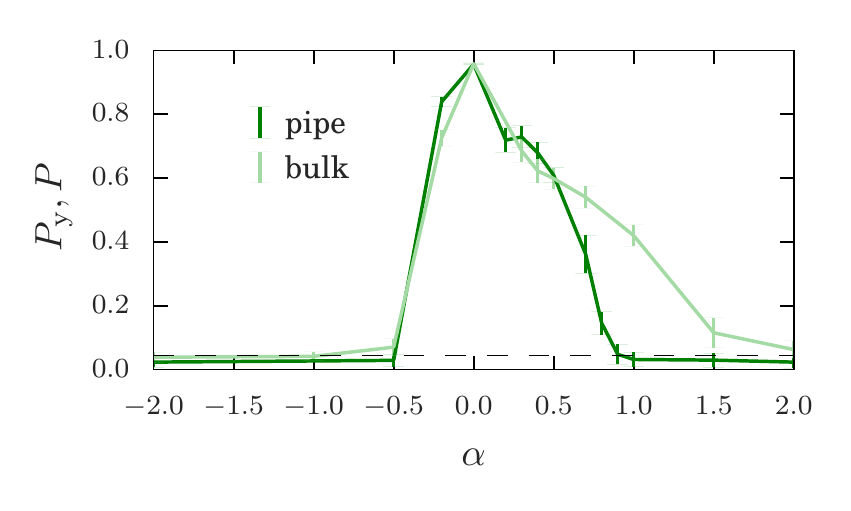}
  \caption{\label{fig:bulk_pipe}
    The polar order parameter $P_{\rm y}$ in a pipe with diameter
    $8\sigma$ as a function of the swimming parameter $\alpha$.
    The polar order $P$ in bulk systems is also shown as a light-colored line.
    The volume fraction is $\varphi\approx 0.25$ in both cases.
    The dashed line stands for the value of $P_0$.
  }
\end{figure}

We conducted simulations over a wide range of combinations
of the swimming parameter $\alpha$ and the volume fraction $\phi$ in the pipe.
From the particle trajectories obtained from these simulations, we
then calculated the polar order parameter along the $y$ direction
(the direction of the pipe elongation),
\begin{align}
  P_{\rm y} =  \left< \cfrac{1}{N_{\rm p}}  \left|\sum_i^{N_{\rm p}}
  e_i^{\rm y} \right|\right> ,
  \label{eq:pipe_Q}
\end{align}
where $e_i^{\rm y}$ is the $y$-component of the swimming direction
$\hat{\boldsymbol{e}}$ of particle $i$ and the angular brackets denote
a time average.
In Fig.~\ref{fig:bulk_pipe}, the values of the in-pipe polar order $P_{\rm y}$
are shown as a function of $\alpha$.
Here, the values of the polar order parameter $P$
in bulk are also shown using a lighter-colored line. 
The volume fraction is $\varphi\approx0.25$
for both cases ($N_{\rm p}=500$ in the pipe, $N_{\rm p}=550$ in bulk).
In Fig.~\ref{fig:bulk_pipe}, we see a good match between
results in bulk and those in pipe over a wide range of $\alpha$.
However, for $0.5<\alpha\le 1.5$, we see clear differences.
In pipes, over this $\alpha$ range, the values of $P_{\rm y}$ are
reduced compared to the bulk results, 
with the gap increasing with an increase in the absolute value of
$\alpha$. Finally, at $\alpha\simeq 0.9$, the dynamics show
a qualitatively different behavior: in the pipe, 
$P_{\rm y}=P_0$ which means the order is completely lost.

\begin{figure}
  \includegraphics[width=\linewidth]{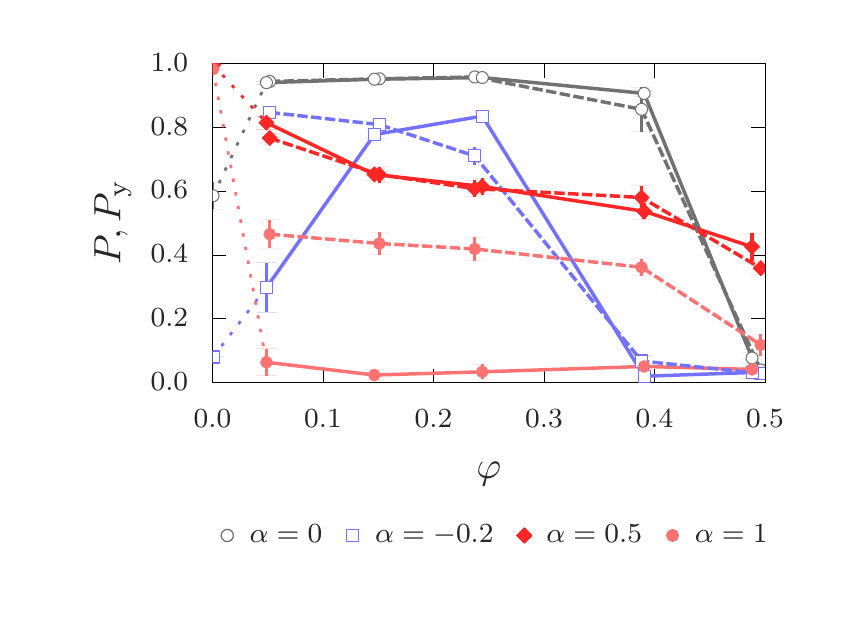}
\caption{\label{fig:order_pipe_big}
  The polar order parameter $P_y$ in a pipe with diameter $8\sigma$
  as a function of volume fraction $\varphi$ and swimming parameter $\alpha$.
  The polar order $P$ in bulk systems is also shown using dashed lines.
  The dotted lines connect the results from multi-particle systems and those from single-particle systems in pipes.
}
\end{figure}

\begin{figure*}
  \begin{center}
\includegraphics[width=\textwidth]{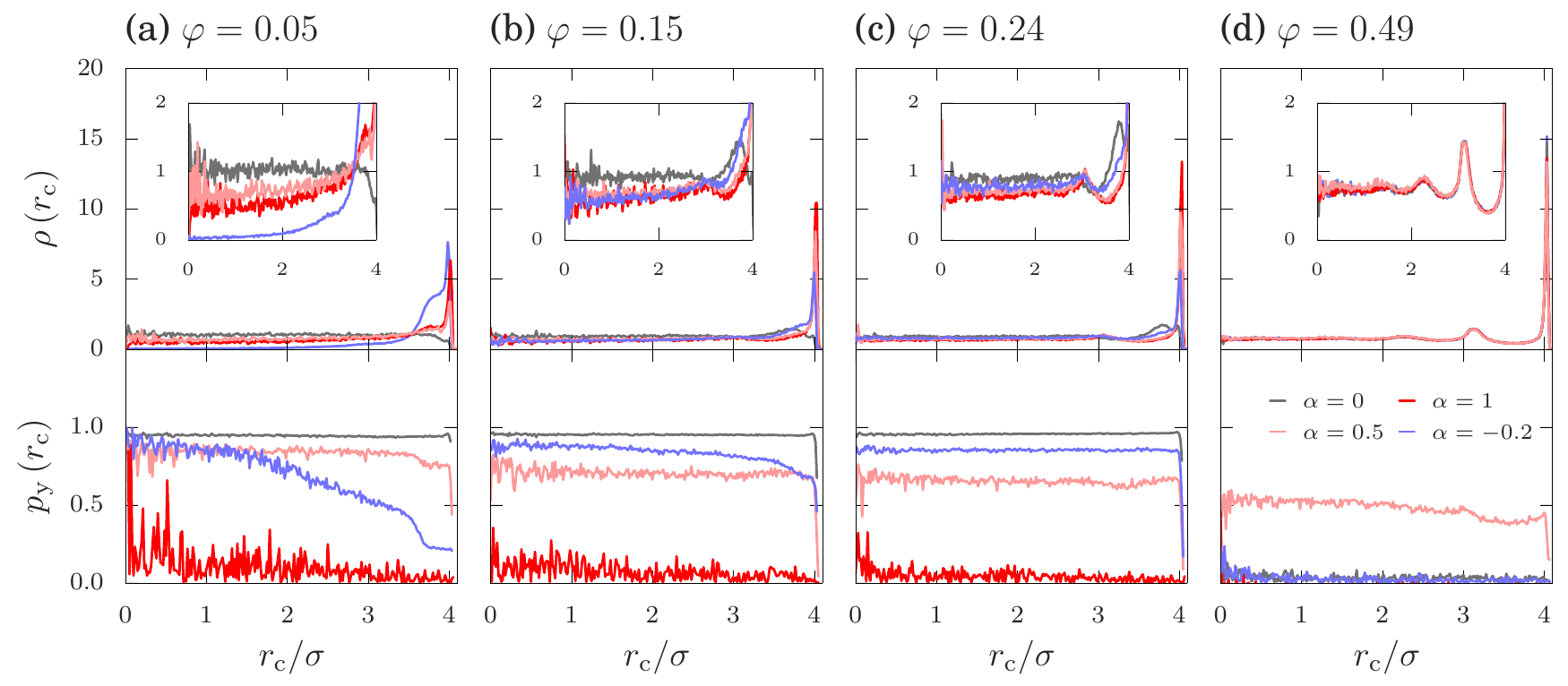}
\caption{\label{fig:dis_pipe}
  The single-body density distribution $\rho$ and the local polar
  order $p_y$ for various combinations of $\varphi$ and $\alpha$.
  Insets show the images magnified in the $y$ direction.
  The horizontal axis represents the distance from the pipe axis normalized by
  the particle diameter.
}
\end{center}
\end{figure*}

In Fig.~\ref{fig:order_pipe_big}, detailed results representing the volume fraction
dependency are shown for four different values of the swimming parameter $\alpha$, which
exhibit qualitatively different behavior, $\alpha=-0.2,0,0.5,1$.
For the volume fraction, we consider five different values in the range of 
$0.05 \le \varphi \le 0.5$ (corresponding to $100\le N_{\rm p}\le 1150$).
As before, the bulk results are shown as dashed lines in Fig.~\ref{fig:order_pipe_big}.
Although the results in the pipe are qualitatively
the same as those in bulk for some of the cases,
several differences can be perceived.
In the case with $\alpha=-0.2$, the value of $P$ at the smallest volume
fraction ($\varphi \approx 0.05$)
shows an obvious reduction in the pipe compared to the bulk.
In addition, for this case ($\alpha=-0.2$),
we observe a rise in $P_y$ at intermediate
volume fraction ($\varphi \approx 0.25$).
In the case with $\alpha=1$, we see no polar order
($P_y\simeq 0$) at any of the volume fractions considered.
Lastly, we note that in the cases of $\alpha=0, 0.5$ we observed
a very good agreement between the results in bulk and those in the
pipe over the whole range of volume fractions.
We also show the value of the polar order $P$
  for the single-particle system in the pipe
  as the data for the infinite dilution systems
  with dotted line
  (though it leads to the volume fraction of $\varphi \approx 0.0005$).
  Note that the value of $P$ for a single neutral swimmer depends on the initial
  orientation: if the initial orientation is parallel to the pipe axis,
  it shows a wave-like trajectory and the value of $P$ is not constant.
  Here, we show the value for the case with a spiral motion where $P$
  is almost constant with time.
  Though we can smoothly extrapolate the missing data in the range below $\varphi=0.05$ for the cases with $\alpha=-0.2,0.5$,
  we see clear jumps for $\alpha=0,1$.
  Because the relaxation time to reach the steady state becomes very long
  as the volume fraction becomes small in the pipe,
  it is very difficult to investigate what is happening at such dilute systems.
  Especially, the clarification of the onset of  jumps of the polar order $P$ for the cases with
  $\alpha=0,1$, or the transition from the single particle motion to
  the collective motion, still remains an open question.

In order to understand these differences from the view point of the
structures inside the pipe, we measured the following two functions,
the local density $\rho\left( r_{\rm c} \right)$ and the local polar order
along the $y$-axis $p_y\left(r_{\rm c} \right)$.
These are defined as a function of the perpendicular distance $r_{\rm c}$ from the symmetry axis of the pipe:
\begin{align}
  \rho\left( r_{\rm c} \right) &= \left< \frac{N_{\rm c}\left( r_{\rm
      c} \right)}{2\rho_0 \pi r_{\rm c} 
    L \Delta r_{\rm c}}\right>,\\
  p_{\rm y}\left( r_{\rm c} \right) &= \left< \frac{1}{N_{\rm c}\left( r_{\rm c}\right)}
  \left| \sum_{i \in \text{bin}}e_i^{\rm y} \right| \right>\label{eq:qinpipe},
\end{align}
where $N_{\rm c}$ is the number of
particles at a distance $r_c$ from the pipe axis,
$\rho_0=\frac{N_{\rm p}}{V_{\rm M}}$
the average number density of the whole system, $\Delta r_{\rm c}$ the
width of the cylindrical bins, and the summation in Eq.~(\ref{eq:qinpipe}) is taken over
all the particles in a given bin.
The local density of the shell matches the overall average value when $\rho
\left( r_{\rm c}\right)=1$.
The results are shown in Fig.~\ref{fig:dis_pipe} for the same values of $\alpha$
shown in Fig.~\ref{fig:order_pipe_big}.
From this figure, we can explain the differences seen in Fig.~\ref{fig:order_pipe_big}.
In the case of $\alpha=-0.2$,
when the volume fraction is small ($\varphi=0.05$, (a)),
particles show a strong tendency to accumulate in the vicinity of the wall and almost
no particles are observed within the central region ($r_c\simeq 0$).
Such a wall accumulation effect has been reported for active
systems in general\cite{Elgeti:2013js,Costanzo:2014iz}, and several
works on squirmers  have also shown this behavior\cite{Oyama:2016eh,Li:2014bu}
At the same time, we can see that the local polar order is reduced
in the vicinity of the wall.
In other words, the pipe can be divided into two distinct regions
from the view point of $p_{\rm y}$: an inner region where the local
order has a value similar to that of the corresponding bulk dispersion,
and an outer region (near to the walls) where the order is reduced.
In the following, we specify the characteristic size of this outer region by $l_{\rm
  c}$, the distance from the wall to the point where the inner region starts.
Because, in the case with $\alpha--0.2$ and $\varphi=0.05$, most particles accumulate in this outer region, where
the order is reduced,
the overall polar order is smaller than that in bulk.
Then, as $\varphi$ is increased,
the particles will become more uniformly distributed, and $l_{\rm c}$
is reduced.
At the same time, due to the wall accumulation effect,
the effective volume fraction in the inner region is smaller
than the global volume fraction.
Taking into account the fact that the polar order increases with
decreasing volume fraction,
this reduction in the volume fraction leads to a higher
local polar order at the inner region, compared to the bulk value at the
same global volume fraction.
As a result of these two effects, namely,
the decrease in $l_{\rm c}$
and the increase of the local order due to the diluteness
at the center, at $\varphi=0.24$,
the value of $P_{\rm y}$ is even larger than that in bulk for the case with $\alpha=-0.2$.
In the case of other values of the swimming parameter ($\alpha=0,0.5$),
such an overshoot is not observed.
This is because for $\alpha=0,0.5$, the volume fraction dependency of the polar
order is so small that the order is determined only by the
value of $l_c$.
For $\alpha=1$,
the value of $l_c$ becomes very large and
we don't see any ordering throughout the pipe, although
the system with $\alpha=1$ does show a nonzero
value for polar order parameter $P$ in bulk.
In the limit when the diameter of the pipe goes to infinity, we expect
the dynamics should converge to that observed in bulk.
Therefore, we should see a qualitative transition at a critical diameter.
This will be investigated in more detail in Sec.~III.~C..
We also note that the behavior of $\rho$ for $\alpha=1$
is very similar to that of $\alpha=0.5$ at all volume
fractions,
although the order formation tendencies are completely different.
At the highest volume fraction $\varphi=0.49$,
because of the screening of the hydrodynamic interactions,
the excluded volume effects become dominant in determining
the structure.
Consequently, the structure is the same for all values of $\alpha$.
However, it is interesting to note
that only in the case of $\alpha=0.5$, do we observe a nonzero
value of $P_{\rm y}$.

We have clarified that in  many particle systems,
the wall effects can be characterized by the accumulation of particles
over a distance of $l_{\rm c}$ close to the walls.
These can lead to both enhancement and inhibition of the
ordering, depending on the degree of accumulation and the size of
$l_{\rm c}$.
While for some cases we do observe a rectification effect due
to the existence of walls,
the origin for such effects seems to have nothing to do with the
causes responsible for the single particle trajectories presented above.

\subsection{Pipe size dependency}
Here, we show results for the pipe diameter dependence of the
dynamics.
First, we present the results for the single-particle systems in a
pipe with a ``small'' diameter.
We present the results for a pipe with $D=3\sigma$ as a
reference system for the small pipe.
For the single-particle systems, we have observed the
same qualitative results as those for $D=8\sigma$ (which we refer to
as the ``big'' pipe).
The only difference that can be seen is the value of $\alpha$ at which the dynamic mode transition occurs.
Interestingly, such a change in the threshold is only seen for a pusher: for the small pipe, the threshold is around $\alpha= -2$.
For a puller, the threshold is the same as the one in the pipe with $D=8\sigma$ and it is  $\alpha=0.4$.
Such asymmetric dependency on the swimming parameter $\alpha$ of the dynamics is also seen
in the polar order parameter $P$ (or $P_{\rm y}$) as a function of $\alpha$, as shown in Fig.~\ref{fig:bulk_pipe}.
To conclude, the single-particle motion does not depend on the pipe size qualitatively.

\begin{figure}
\includegraphics[width=\linewidth]{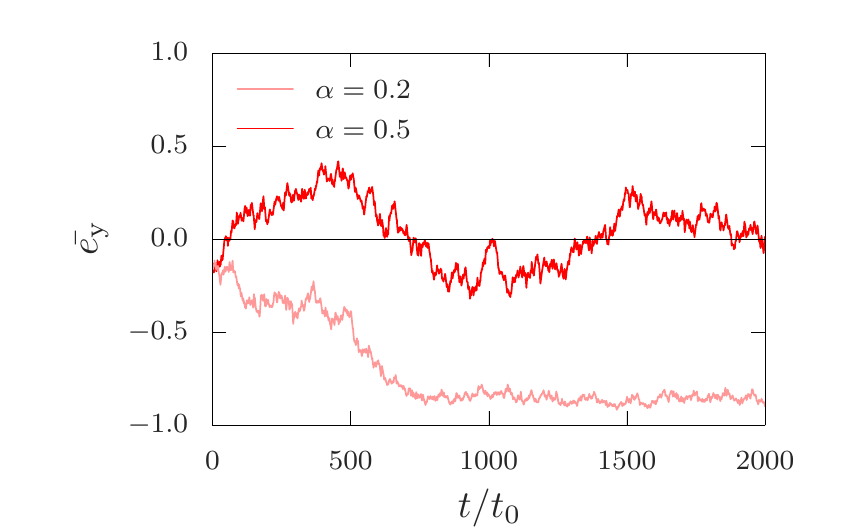}
\caption{\label{fig:t_e}
  The time evolution of the temporal order $\bar{e^{\rm y}}$ for 
  systems with $\alpha=0.2,0.5$ in the small pipe with $D=3\sigma$ at
  $\varphi=0.28$.
  The horizontal axis gives the time normalized by
  $t_0=\sigma/U_0$, where $U_0=\frac{2}{3}B_1$ is the steady state velocity of an
  isolated swimmer.
  The vertical axis represents the value of temporal order.
}
\end{figure}

\begin{figure}
\includegraphics[width=\linewidth]{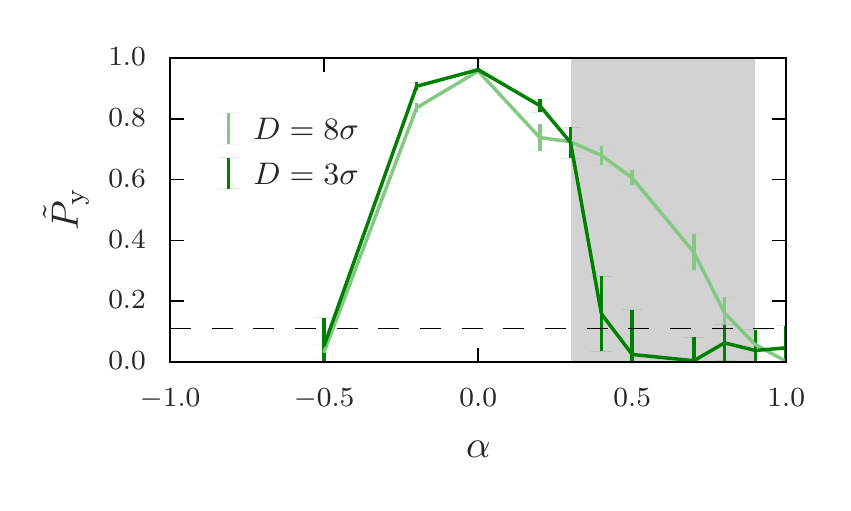}
\caption{\label{fig:order_pipe_small}
  The dependency on $\alpha$ of the polar order $\tilde{P}_{\rm y}$ in pipes.
  The dark lines are for the pipe with
  $D=3\sigma$, while the light lines are for $D=8\sigma$.
  The volume fraction is $\varphi\approx 0.25$ for both cases.
  The shaded region represents the region where the anomalous behavior
  was reported in ref.\cite{Oyama:2016eh,Oyama:2016vj}.
}
\end{figure}

Next, we investigated the polar order of many particle systems in the small pipe.
In the small pipe, over a specific range of parameters, the value of
the polar order doesn't show a stable nonzero value.
In Fig.~\ref{fig:t_e}, the time evolution of the 
temporal order $\bar{e}^{\rm y}$, defined in
Eq.~(\ref{eq:ave_e}), is given for two cases to illustrate both the
steady and unsteady behaviors.
\begin{align}
  \bar{e}^{\rm y}=\cfrac{1}{N_{\rm p}} \sum_i^{N_{\rm
        p}}e_i^{\rm y}.
  \label{eq:ave_e}
\end{align}
Both results are for a system with $D=3\sigma$ and
$\varphi=0.28~(N_{\rm p}=80)$.
The steady results are for
$\alpha=0.2$, the unsteady one for $\alpha=0.5$.
As seen here, in the unsteady system, even the sign
of $\bar{e}^{\rm y}$, which represents the direction of motion,
changes in time; while in the
systems with stable order, such
change is never observed.
This change of direction shows the absence of any long-time
persistent order in the pipe.
In order to evaluate such direction-flipping behavior,
we define an alternative in-pipe polar order parameter that can also indicate the
persistence of the motion:
\begin{align}
  \tilde{P}_{\rm y} =   \left|\left< \cfrac{1}{N_{\rm p}} \sum_i^{N_{\rm p}}
  e_i^{\rm y} \right> \right|.
  \label{eq:persistency}
\end{align}
Only the order of taking the average and the absolute
value is changed with respect to the standard definition of the polar
order $P_{\rm y}$, Eq.~(\ref{eq:pipe_Q}).
For the cases in which the direction doesn't change,
the values $P_{\rm y}$ and $\tilde{P}_{\rm y}$ are essentially the same.
We measured $\tilde{P}_{\rm y}$ in the small pipe for various values of
$\alpha$ and compared the results with those obtained for the big pipe.
The results are shown in Fig.~\ref{fig:order_pipe_small} at a volume
fraction of $\varphi \approx 0.25$.
We observe a remarkable differences over the range
$0.4\le\alpha\le 0.7$.
In the small pipe for this range of values of $\alpha$,
$\bar{e}^{\rm y}$ has no steady state value and changes
directions, as described above.
Actually, for such systems, the values of $\tilde{P}_{\rm y}$
are around $P_0$, which means that the
system has no order.
We consider this collapse of the order as an
order/disorder phase transition.
The gray-shaded region in Fig.~\ref{fig:order_pipe_small} indicates the
range of $\alpha$ at which anomalous behaviors were reported in
preceding studies\cite{Oyama:2016eh,Oyama:2016vj}.
In ref.\cite{Oyama:2016eh}, the dynamics of swimmer dispersions
between flat parallel walls was studied.
It was reported that, over this gray-shaded range of $\alpha$,
particles form big clusters, which exhibit a traveling wave-like motion,
bouncing back and forth between walls.
Interestingly, the shaded area corresponds very well to the region
over which we observe the pipe-size
dependent order/disorder transition in the present work.
We refer to pullers in these regions as  ``intermediate'' pullers.
For $\alpha$ values other than those of intermediate pullers,
the results are roughly the same between $D=8\sigma$ and $D=3\sigma$.
We note that we have confirmed that the same qualitative results are
obtained for different values of $\varphi$, at least
over the range we have considered~($0.05 \le \varphi \le 0.25$).
We would also like to stress that in bulk, we have observed a similar
order/disorder phase transition when we change the volume
fraction\cite{Evans:2011cw,Oyama:2016vj}:
when the volume fraction becomes very high, none of the swimmers except
intermediate pullers are able to maintain the polar order.
In contrast, in pipes, only intermediate pullers lose
their polar order when the size of the pipe becomes small.

\begin{figure}
\includegraphics[width=\linewidth]{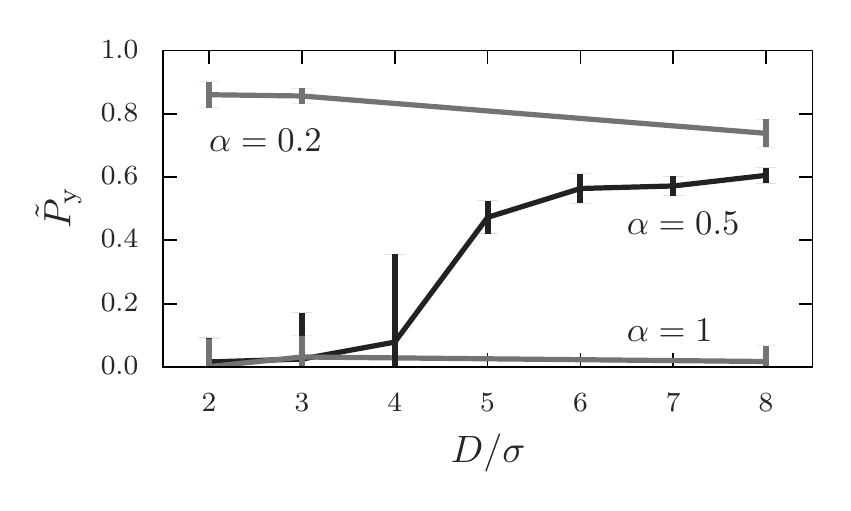}
\caption{\label{fig:size_d_05}
  The pipe diameter dependency of the in-pipe polar order $\tilde{P}_{\rm y}$.
  The horizontal axis represents the pipe diameter
  $D$ normalized by the particle diameter.
  The volume fraction is $\varphi\approx 0.25$ for all cases.
}
\end{figure}

We have also investigated the detailed size dependency of the polar order for intermediate pullers.
In Fig.~\ref{fig:size_d_05}, the values of the polar order $\tilde{P}_{\rm
  y}$ are shown as a function of the pipe diameter.
Here, the results are shown not only for intermediate pullers $\alpha=0.5$,
but also for weak and strong pullers $\alpha=0.2, 1$, which don't show
the phase transition.
The volume fraction is $\varphi\approx 0.25$ for all cases ($40\le
N_{\rm p} \le 500$, depending on the size of the pipe).
We see an abrupt drop of $\tilde{P}_{\rm y}$ at the critical diameter
$D_{\rm c}=4\sigma$,
though at higher $D$, the values are constant.
In systems with $D \le 4\sigma$,
the value of $\tilde{P}_{\rm y}$ does not achieve the steady state,
similar to the situations seen in
Fig.~\ref{fig:order_pipe_small}.
In other words, in such small pipes, intermediate pullers cannot maintain the
polar order anymore, though temporally they can exhibit order.
We believe this sensitivity to the pipe size suggests that
intermediate pullers need clusters bigger than a characteristic size $D_{\rm c}$
to stabilize the polar order.

\begin{figure*}
  \begin{center}
\includegraphics[width=\textwidth]{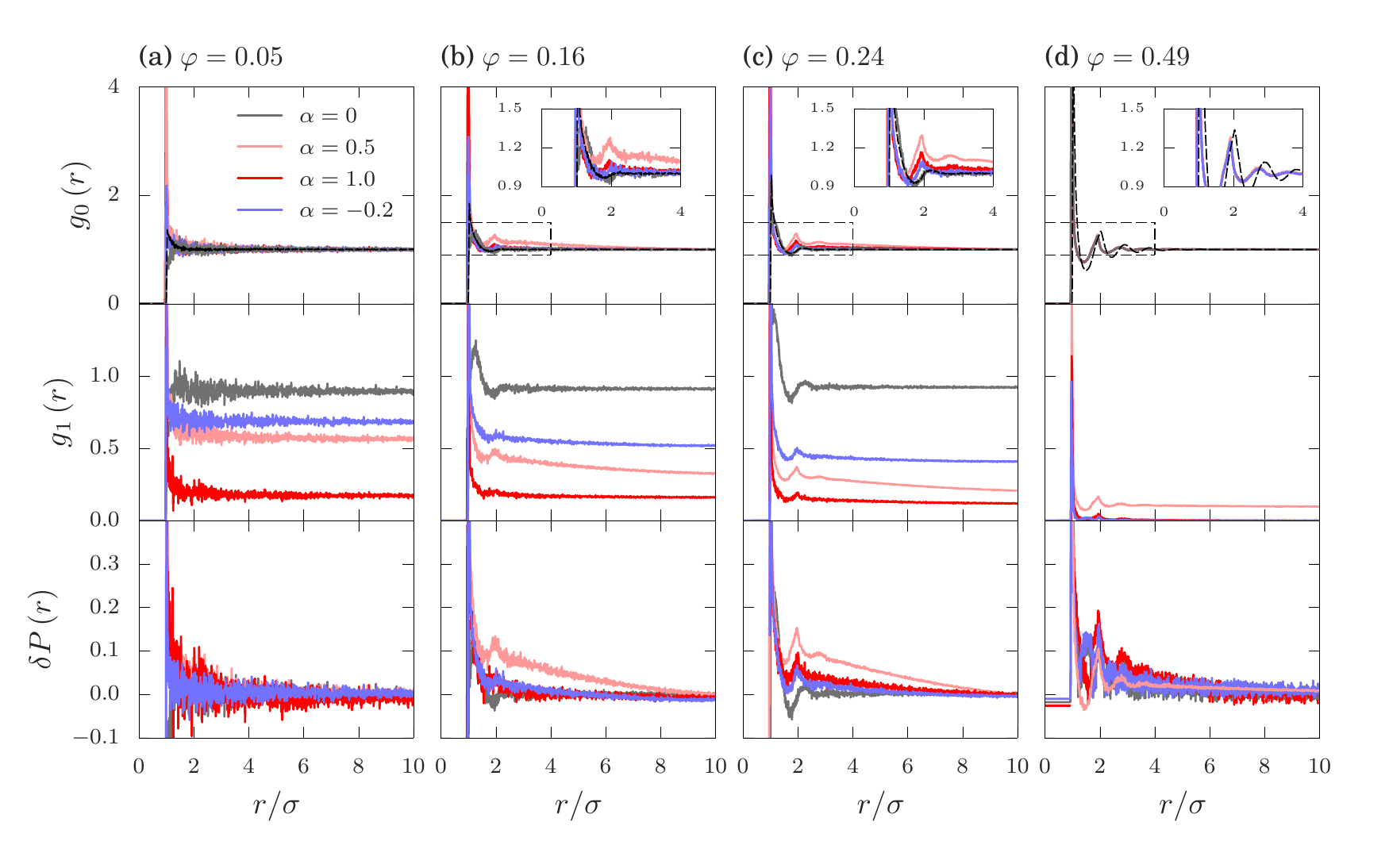}
\caption{\label{fig:gen_dis}
  The generalized radial distribution functions $g_0$, $g_1$ and
  the local deviation of the order $\delta P$ with respect to bulk values
  for various combinations of the volume fraction $\varphi$ and the swimming parameter $\alpha$.
  Insets show the images magnified in both $x$ and $y$ directions.
  The horizontal axis represents the distance from the reference
  particle normalized by the particle diameter.
}
\end{center}
\end{figure*}

In order to specify the size of the cluster in bulk,
we performed bulk simulations and measured
the generalized radial distribution functions
$g_n\left( r \right)$\cite{Alarcon:2013dg},
\begin{align}
  g_n\left( r \right) = \left< \frac{1}{4\pi r^2 \Delta
    r}\frac{1}{\rho_0 \left( N_p-1\right)} \sum_{i=1}^{N_p}\sum_{j\in
    \text{bin}} P_n\left(
  \text{cos}\theta_{ij}\right) \right>\label{eq:gen_dis},
\end{align}
where $r$ represents the distance from a reference particle $i$ to a
second particle $j$,
$\theta_{ij}=\text{cos}^{-1}\left( \hat{\boldsymbol{e}}_i\cdot
\hat{\boldsymbol{e}}_j\right)$ the relative angle between the swimming
directions of both particles, and $P_n$ is the $n$-th degree Legendre
polynomial.
The zeroth-order function $g_0\left( r \right)$ is the standard radial
distribution function, and the first-order function $g_1\left( r
\right)$ represents the local degree of orientational order.
In the limit when $r\to \infty$, we expect
$g_0\left( r\to \infty \right) \sim 1$ and $g_1\left( r\to \infty
\right) \sim P^2$.
Therefore, we can define
the excess local order $\delta P \left( r \right)=
\sqrt{g_1}-P$, in such a way that it will converge to zero in the limit of $r\to \infty$.
The bulk system considered here has a cubic domain with a linear dimension of
$128\Delta \approx 20 \sigma$.
Periodic boundary conditions are used in all directions, and the
particles have the same diameter and interface thickness as in the
simulations discussed above.
The volume fraction is tuned by changing the number of particles,
$1000\le N_{\rm p} \le 9000$.
The calculated results for $g_0\left( r \right)$, $g_1\left( r \right)$
and $\delta P \left( r \right)$ are shown in
Fig.~\ref{fig:gen_dis}, for $\varphi=0.05, 0.16, 0.24, 0.49$ and
$\alpha=-0.2,0,0.5,1$, corresponding to the confined systems in Fig.~\ref{fig:dis_pipe}.
In addition, results for $g_0\left( r \right)$ for passive colloidal
systems are shown to aid in comparison. For these passive systems, the temperature is set so that $k_{\rm B}T=\epsilon$,
where $\epsilon=\frac{\Delta\eta^2}{\rho_{\rm f}}$ is the energy
unit in our system.
The exact value of the temperature does not affect the
radial distribution function, provided that it is small enough.

To begin the analysis of the structure, we discuss the behavior of $g_0\left( r\right)$ for
the reference system of passive colloids.
At the smallest volume fraction~($\varphi=0.05$),
it shows only a small first peak around $r=\sigma$,
and quickly decays, converging to one around $r=1.5\sigma$.
The intensity of this first peak becomes larger as the volume
fraction increases.
At a volume fraction of $\varphi=0.16$,
we observe a shallow undershoot after the first peak.
The correlation length becomes longer and the value converges to one at
around $r=2\sigma$.
At $\varphi=0.24$,
it starts to show a second peak at a position $r\gtrsim 2\sigma$, though the intensity is very small.
Finally, at $\varphi=0.49$,
the system is in a mixed state of ordered and disordered
phases\cite{Russel:1992wr}, and shows four clearly distinguishable
peaks~(see inset magnified figures in Fig.~\ref{fig:gen_dis}(d)).

Now, we would like to present the results for the active systems.
In the case of neutral swimmers ($\alpha=0$),
for all the volume fractions but the highest one,
the first peak is smaller or shifted to larger distances compared to
the passive system.
In addition, the first peak doesn't even appear at the lowest volume
fraction, $\varphi=0.05$.
These behaviors may indicates that there is an effective repulsive interaction
between particles caused by the flow field.
Due to this repelling flow field, neutral swimmers tend to get
distributed uniformly.
Similar effects can be seen in the pipe, as shown in
Fig.~\ref{fig:dis_pipe} where neutrals swimmers show a much more
uniform distribution compared to the other types of swimmers.
The correlations beyond the first peak follow closely those of the passive system, while other swimmers show longer ranged correlations.
This tendency may reflect the fact that the flow field around a neutral swimmer
(decaying as $r^{-3}$) is more localized, because of the lack of the force
dipole term (decaying as $r^{-2}$; the second term in Eq.~\ref{eq:Sq_2}).
Regarding $g_1\left( r\right)$, the qualitative behavior follows
that of $g_0\left( r\right)$.
We note that the neutral swimmers exhibit a unique behavior around
$r=2\sigma$, where $\delta P \left( r\right)$ shows negative values for
intermediate values of $\varphi=0.16,0.24$.

In the case of $\alpha=-0.2$ and $1$,
the results are quite similar for all the volume fractions,
although the peak heights show slight differences.
In these cases, the particles seem to show effective attraction due
to the flow field, resulting in higher peaks for $g_0\left( r \right)$.
As before, $g_1\left( r\right)$ follows the qualitative behavior seen in
$g_0\left( r \right)$.
Besides the higher peaks,
the correlation length is longer than that seen for neutral swimmers.
However, the bulk values are reached by $r\approx 6\sigma$ in all cases.

In the case with $\alpha=0.5$, or intermediate pullers, many unique behaviors can be
observed.
First of all,
$g_0\left( r\right)$ shows much more
outstanding peaks than in any of the other cases.
At $\varphi=0.16$, only intermediate pullers show a definite second peak,
and at $\varphi=0.24$, the second peak is much higher than in the
other systems, and a third peak even appears.
A similar peak behavior can be observed also in the plot of $\delta
P\left( r\right)$.
In addition, we see a very long tail, which is not seen in the other cases.
All these unique behaviors for intermediate pullers can be
understood as indirect evidence for
their already known dynamic clustering tendency\cite{Oyama:2016eh,Alarcon:2013dg}.
The high peaks reflect the clustering behavior,
and the long tail in $g_1\left( r \right)$ shows the highly
localized ordering.
The nonzero order at high volume fractions helps explain
how intermediate pullers can swim collectively, even under such
extreme conditions.
We note that the critical diameter of the pipe $D_{\rm c}$ is
bigger than any of the peaks detected here.

At the highest volume fraction, $\varphi=0.49$,
the shapes of $g_0\left( r \right)$ 
are the same for all $\alpha$,
while only the passive system shows a different shape.
Regarding the neutral swimmer systems, $\alpha=0$, this is notable
because this system shows quite similar shapes at all the other volume fractions.
The shape difference of $g_0\left( r\right)$
between the passive and active systems at $\varphi=0.49$
may be due to the phase transition of the passive system.
For passive hard-spheres,
the volume fraction in the range $0.5<\varphi<0.55$ is known
as the transient state of the ordered state and
the disordered state\cite{Russel:1992wr}.
In the passive system at $\varphi=0.49$, shown in Fig.~\ref{fig:gen_dis},
such an ordered state seems to appear,
while the active systems are still completely in the fluid phase.
At this volume fraction, because the
excluded volume effect is the dominant factor to determine the
structure, rather than the flow field, all the active systems show the same shape for $g_0$.
In summary, although we could not obtain a direct connection between the bulk
structural information and the in-pipe order/disorder phase transition,
we believe our results for the structure present indirect evidence for the importance of
the dynamic clustering on the stabilization of the polar order for
intermediate pullers systems.
In order to verify the exact connection between these two quantities,
we have to conduct a more detailed analysis on the dynamic clustering in bulk.
  We note that if we use a finer resolution of swimmers,
  we see the decrease in 
  the exact value of the critical pipe diameter $D_{\rm c}$ for the intermediate pullers:
  if we double the particle size, $\sigma=12\Delta$,
  we obtain the critical diameter $D_{\rm c}=2\sigma$.
  (changes seems to be able to be seen only for the value of $D_{\rm c}$).
  In other words, for intemediate pullers under strong confinement
  we require higher resolution to accurately resolve the flow around the particles.
  This resolution dependence can imply that
  smaller size clusters are enough for the polar order formation,
  like the ones with size corresponds to the peaks in $g\left( r\right)$.
  However, with the contemporary calculation power,
  the bulk simulation with such a finer resolution is too expensive
  to conduct with a realistic time and we cannot make a fair comparison between the results
  from the bulk simulation.
  Therefore, the discussion on the precise cluster size remains an open question.
  We confirmed the existence of the phase transition
  and the results are consistent between the two systems with different particle resolution.
Our previous results \cite{Oyama:2016vj} support this view on the
importance of dynamic clustering.
In ref.\cite{Oyama:2016vj}, we investigated whether or not the polar order seen in bulk 
can be explained only by the repetition of binary collisions.
We found that this is the case, except for systems of intermediate
pullers, which means that many particle interactions might be
important in systems of intermediate pullers.

Finally, we have also considered pipes with diameters larger than
$8\sigma$.
As the pipe size increases, we can expect that the
dynamics approaches that seen in bulk systems.
Indeed, for systems with $\alpha\ne 1$, we see in
Fig.~\ref{fig:order_pipe_big} that the order parameters in confinement
agree very well with the bulk values. Thus, for these values of $\alpha$, $D=8\sigma$ seem to be big enough.
Therefore, we have only considered larger diameters for
$\alpha=1$, to investigate whether a nonzero value of the polar order is
recovered for a large enough pipe diameter.
We considered diameters up to $D=18\sigma$ but were not able to
recover the bulk values in this case.
Thus, the hypothesis stated in section III.~B. is still not solved
within the range of parameters considered in the present study.
Although it is possible that the dynamics can change for even larger
diameters, we will not discuss this problem any more, 
as it lies outside the scope of the present  work.

\subsection{Confinement shape dependency}
\begin{figure}
\includegraphics[width=\linewidth]{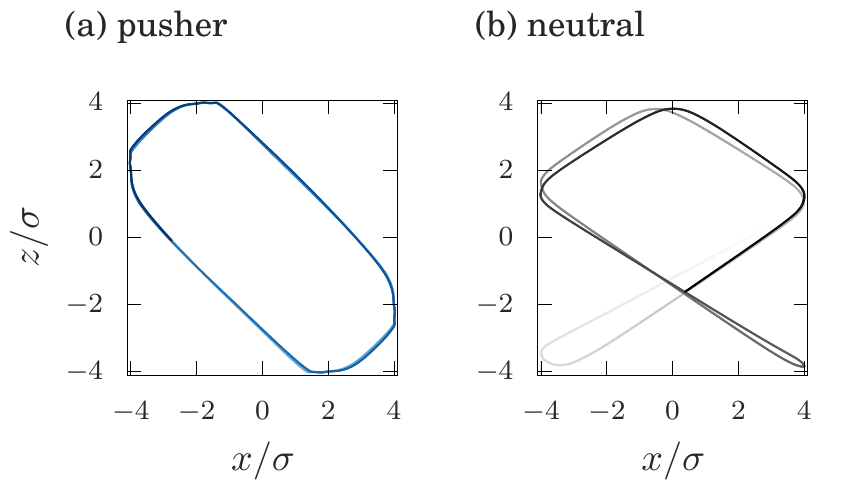}
\caption{\label{fig:pro_tra_duct}
  The projection of the trajectories of a swimmer in duct onto the
  $xz$-plane.
  (a) Pusher with $\alpha=1$. (b) Neutral swimmer.
  The lines represent the projections of the
  trajectories, with color intensity used to represent the time.
}
\end{figure}

The dynamics of fluids in pipes is known
to depend on the shape of the cross section.
Therefore, we have also investigated the effect of the shape
dependency on the dynamics of the swimmers.
So far, we have only considered dynamics in
cylindrical pipes, with
circular cross sections.
Here, we consider pipes with a rectangular cross section.
We refer to such confinement geometries as ``ducts''.
We considered two sizes of ducts, specified by a lateral size of
$3\sigma$ and $8\sigma$, respectively.
Regarding the single-particle dynamics,
the trajectories are qualitatively different from those in pipes
for the case of pushers and neutral swimmers.
Pullers will swim parallel to the duct axis, as they do in pipes.
In ducts, regardless of the size,
the trajectories of the pushers are spiral.
Of course, reflecting the duct shape,
the projections onto the $xz$-plane are no longer circular (Fig.~\ref{fig:pro_tra_duct}).
The neutral swimmer shows a similar spiral trajectory, but it has a strong
dependency on the initial state and sometimes the trajectories can become
more complicated (Fig.~\ref{fig:pro_tra_duct}).
Considering all the results for the single-particle system of pushers,
we can conclude that the emergence of the circular closed orbits
depends strongly on the curvature of the confining walls.
Therefore, if we increase the diameter of the pipe continuously from $8\sigma$,
at some point we will find a critical value for pushers to lose their
orbit-type characteristics.
Such a detailed study is beyond our scope.

With regards to the many particle dynamics, we don't see any big differences from
those in pipes.
Namely, the results are almost the same as those in bulk when the duct
size is big enough,
and the ordering seen for intermediate pullers collapses when the size is decreased.
As a result, intriguingly, from the view point of the polar order of the
many particle dispersion, the results are qualitatively the same between pipes and ducts.
However, it is possible that we can observe unique dynamics in a skewed
confinement.

\section{\label{sec:level5}Conclusion}
We investigated the polar order formation in pipes,
conducting three-dimensional direct numerical simulation with fully resolved hydrodynamics.
As a result of the investigation over a broad range of parameters,
we confirmed that in most cases the dynamics of many particle
systems in pipes match very well those in bulk. We have only observed
considerable differences over a small range of parameters.
We have presented an explanation  for the emergence
of wall effects by measuring the structural information.
They are mostly due to the wall accumulating effects and the reduction
in the degree of order in the vicinity of the walls.
Also, we have studied the effects of the change in pipe size.
We have clarified that for a specific range of the swimming parameter,
the system undergoes an order/disorder phase transition when the size of
the pipe shrinks.
Considering the results of the structural analysis 
in bulk,
we believe that this is due to the intrinsic clustering tendency of
the swimmers.
There seems to exist a minimum cluster size which is
necessary to maintain the polar order of intermediate pullers.
If we change the shape of the confinement,
we found no qualitative differences, from the view point
of the polar order.
These results can be utilized to address the transport problem of microswimmers, however,
further studies, like in winding pipes or in pipes with abrupt
expansion/contraction are desired for realistic applications.

\section{Acknowledgments}
We thank H. Ito for enlightening discussions.
This work was supported by the Japan Society for the Promotion
of Science (JSPS) KAKENHI Grant No. 26247069 and also by a Grant-in-Aid
for Scientific Research on Innovative Areas
 Dynamical ordering of biomolecular systems for creation of
integrated functions 
(No. 16H00765) from the Ministry of Education, Culture, Sports, Science,
and Technology of Japan.
We also acknowledge the supporting program for
interaction-based initiative team studies
(SPIRITS) of Kyoto University.




\end{document}